\documentclass[aps,prl,twocolumn,showpacs,superscriptaddress]{revtex4}
\usepackage{graphicx}

\begin{document}


\title{Determination of the lowest energy structure of Ag$_8$ from 
first-principles calculations}


\author{M. Pereiro}
\email[Email address: ]{fampl@usc.es}
\affiliation{Departamento de F\'{\i}sica Aplicada, Universidade de Santiago
de Compostela, Santiago de Compostela E-15782, Spain.}
\author{D. Baldomir}
\affiliation{Departamento de F\'{\i}sica Aplicada, Universidade de Santiago
de Compostela, Santiago de Compostela E-15782, Spain.}


\date{\today}

\begin{abstract}
The ground-state electronic and structural properties, and the electronic excitations
of the lowest energy isomers of the Ag$_8$ cluster are calculated using
density functional theory (DFT) and time-dependent DFT (TDDFT) in real time and real space scheme,
respectively. The optical spectra provided by TDDFT predict that the D$_{2d}$ dodecahedron isomer
is the structural minimum of Ag$_8$ cluster. Indeed, it is borne out by the experimental findings.
\end{abstract}


\maketitle



%

In latter years, a new field has emerged from the understanding, control and 
manipulation of objects at nanoscale level (nano-objects). It is commonly known as 
nanoscience. This field involves physics, chemistry and even engineering, and
addresses a huge number of important issues starting from basic science
and ending in a large variety of technological applications \cite{baletto}.
Among the nano-objects, the small clusters or nanoclusters play a very important
role, since they are the bricks of nanoscience. Therefore, the study of small
clusters deserves a special attention. In this respect, 
the steps to follow for a complete description of a cluster can 
be summarized in the following three questions:  
what is the lowest energy structure?, what is the effect of increasing or decreasing
the temperature on the structural properties of a cluster? and the last step 
deals with the kinetic effects in the formation of the nanocluster. Here,
we are only concerned about the first question for the silver octamer,
leaving the other two questions open for future investigations. 

From the theoretical point of view, first-principles methods give an enormous advantage 
for understanding, projecting and inventing new materials that is reflected in the huge number
of articles published in the field of materials science. Likewise, density functional theory (DFT) 
has emerged as a new
and promising tool for {\it ab initio} electronic structure calculations and gives valuable information
about the geometry of nanoscale systems \cite{jena} but unfortunately it 
not always predict the correct structure of the
cluster under consideration. In this regard, the silver octamer belongs to the group of the controversial
systems for which the lowest energy structure is unresolved by DFT. 

Recently, P. Radcliffe {\it et al.} \cite{radcliffe} have proposed Ag$_8$ clusters 
embedded in helium droplets as a suitable system for light amplification based on an optically
accessible long-living excited state (E*) and thereby, from the theoretical point of view, the
determination of the structure becomes a key point as the first step for identifying
and controlling the levels that populate E*. Up to now, it has not been 
possible to make a reliable theoretical prediction of the most stable structure of Ag$_8$. 
A review of the literature reveals that there are two competing geometries in eight-atom 
clusters of s-electron elements, having
T$_d$ and D$_{2d}$ symmetry. In fact,
different levels of theories favor different geometries: DFT in its local
density approximation (LDA) \cite{fournier}, multireference configuration interaction 
method \cite{bonacic1}, a tight-binding approach \cite{zhao}, and the many-body perturbation theory-based
calculations \cite{huda} give the
D$_{2d}$ geometrical shape as the lowest energy structure of Ag$_8$ whereas
the equation-of-motion coupled cluster method 
\cite{bonacic}, time-dependent DFT only at LDA level \cite{yabana} and   
molecular-dynamics simulations \cite{erkoc} predict a T$_d$ structure as the structural minimum.
It was reported in Refs.~\cite{bonacic,bonacic1} that 
the D$_{2d}$ geometry is favored energetically over T$_d$ symmetry when explicit correlation treatments for 5s 
electrons are included, but
since the calculated energy difference between T$_d$ and D$_{2d}$ isomers is very small, the
predicted theoretical ordering is uncertain.
One way of solving this vexing problem comes from the hand of the time-dependent density functional
theory (TDDFT) \cite{casida} that is a generalization of traditional ground stationary state DFT to treat
the dynamic response of the charge density to a time-dependent perturbation. TDDFT is 
a powerful methodology towards the calculation of the 
optical spectra, and thereby gives access to excited-state information. 

In this brief report, 
we have calculated the optical response of the T$_d$ and D$_{2d}$ geometries and they were compared both
with each other and the experimental findings. The atomic positions were fully optimized with 
an all-electron DFT
implementation at the generalized gradient approximation (GGA) level, representing an improvement over
other TDDFT studies in small silver clusters \cite{yabana}. This work relies on the combination of the
traditional DFT and its generalization to excited states, as a promising tool for elucidating
structures that DFT by its own is unable to predict. Here, we 
demonstrate that the D$_{2d}$ structure is the structural minimum of Ag$_8$ and the calculated
spectra allow us 
to estimate the interaction of Ag$_8$ with the surrounding helium or argon matrix presented in experimental
observations. 

With the aim of elucidating the lowest energy structure of Ag$_8$ cluster, we have
performed density functional theory-based calculations consisting of 
a linear combination of Gaussian-type orbitals-Kohn-Sham-density-functional
methodology (LCGTO-KSDFM) to obtain the structural and 
electronic ground-state properties \cite{salahub}, and 
a TDDFT implementation to compute the electronic excitations \cite{marques}. 
For the former, all-electron calculations were carried out
with DEMON-KS3P5 \cite{salahub} at GGA level to 
take the exchange-correlation (XC) effects into account \cite{perdew}. An orbital
basis set of contraction pattern (633321/53211*/531+) was used in conjunction
with the corresponding (5,5;5,5) auxiliary basis set for 
describing the s-, p- and d-orbitals \cite{huzinaga}. The grid for numerical
evaluation of the XC terms had 128 radial shells of points and each shell had
26 angular points. Spurious one-center contributions to the XC forces, typically 
found in systems with metal-metal bonds when
using a nonlocal functional, are 
eliminated in a similar way as has been done in Ref.~\cite{versluis}.
Trial geometries were fully optimized without symmetry and
geometry constraints for different multiplicities using the Broyden-Fletcher-
Goldfarb-Shanno algorithm \cite{broyden}. The multiplicities were ranged from 1 to 11 and in all
reported structures the singlet state was favored energetically. During the optimization, the convergence criterion 
for the norm of the energy gradient was fixed to $10^{-4}$ a.u. while it was $10^{-7}$ a.u. for 
the energy and $10^{-6}$ a.u. for the charge density.
For the latter,  after inserting the atomic coordinates of the converged structures provided by DEMON-KS3P5
all the dynamical quantities are computed by evolving the electronic wave functions in real time
and real space \cite{marques}. The electron-ion interaction is described through the
Hartwigsen-Goedecker-Hutter
relativistic separable dual-space gaussian pseudopotentials \cite{hartwigsen} and the XC effect
were treated in the GGA, implemented via the Perdew-Burke-Ernzerhof functional \cite{perdew1}. The
grid in real space to solve the Kohn-Sham equations consists in a sum of spheres around each
atom of radius 5.5 \AA~and a mesh spacing of 0.23 \AA. The time step for the propagation
of the electronic orbitals was fixed to 0.0013 fs, which ensures the stability of time-dependent
propagation. An artificial electronic temperature of 10 K was included according to the
Fermi-Dirac function used to distribute the electrons among the accessible states.

\begin{figure*}
\includegraphics[width=17.8cm,angle=0]{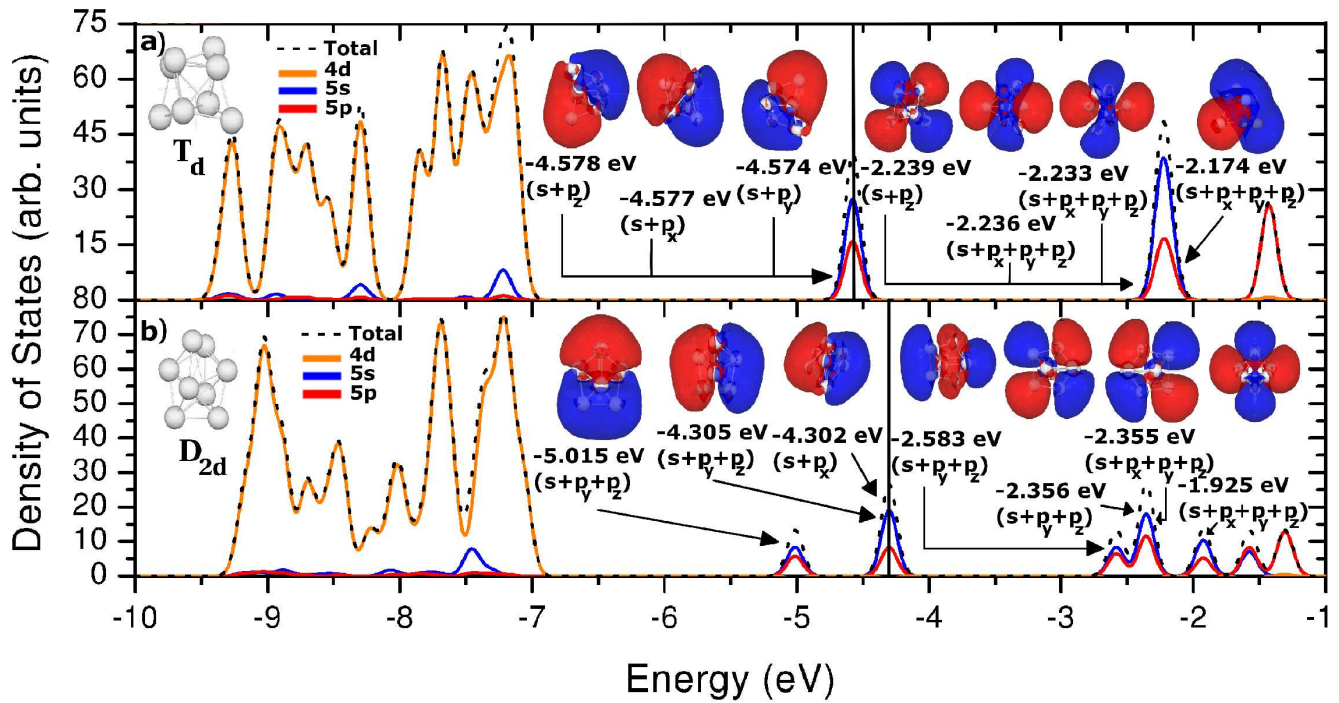}
\caption{\label{fig1}(color online). Energy levels, partial and total density of states and 
shapes of the delocalized molecular orbitals for the higher-lying occupied and
lower-lying unoccupied levels of Ag$_8$ isomers: a) T$_d$-TT
and b) D$_{2d}$-DD. The orbitals below the Fermi level
are double occupied; for
these levels only one molecular orbital is presented. The solid vertical line represents  
the Fermi level.}
\end{figure*}

After a review of the literature on silver clusters \cite{fournier,zhao,bonacic,huda,bonacic1},
we have decided to optimize, as a good candidate to the structural minimum of the octamer,
the following isomers of Ag$_8$: a D$_{2d}$ dodecahedron (D$_{2d}$-DD), which can also be
viewed as a distorted bicapped octahedron, a T$_d$ tetracapped tetrahedron
(T$_d$-TT) and a C$_s$ 1-pentagonal bipyramid (C$_s$-PBP) in Fournier's notation \cite{fournier}.
The main results (density of states (DOS), optimized structures, 
polarizabilities, ground state energies, \ldots) of
the electronic structure calculations are
collected in Table~\ref{table1} and Fig.~\ref{fig1}. The LCGTO-KSDFM calculations clearly show that the
C$_s$-PBP geometry is energetically far from the lowest energy structure by an amount of
181.25 meV. Therefore, we will concentrate our attention in D$_{2d}$ and T$_d$ structures.
\begin{table}
	\caption{\label{table1}Ground-state energies relative to the most stable isomer (D$_{2d}$-DD)
and electronic structure properties of the
DFT-optimized Ag$_8$ cluster isomers. 
The Fermi level is denoted by E$_f$ and $\Delta\xi$ stands for the HOMO-LUMO gap.
The mean static polarizability $\bar{\alpha}$ and
the polarizability anisotropy $\Delta\alpha$ were calculated under
the influence of an external electric field of strength 0.0005 a.u..}
\begin{ruledtabular}  
	\begin{tabular}{ccccccc}
	Symmetry & $\Delta$E$_{DFT}$ & E$_f$ &  $\Delta\xi$ & $\bar{\alpha}$ & $\Delta\alpha$\\
&(meV) & (eV)    & (eV) & (\AA$^3/$atom) & (\AA$^3/$atom)\\
\hline
D$_{2d}$& 0.00 & -4.302 & 1.719 & 6.32 & 1.22\\
T$_d$ & 6.14 & -4.574  & 2.335 & 6.46 & 0.01\\
C$_s$ & 181.25 & -4.048 & 1.327 & 6.45 & 1.71\\
\end{tabular}  
\end{ruledtabular}  
\end{table}

Despite the fact that the lowest energy isomer 
corresponds to a D$_{2d}$ symmetry, it should be noted that the ground-state energy difference
between D$_{2d}$-DD and T$_d$-TT isomers is very small ($\Delta E_{D_{2d}\rightarrow T_d}=6.12 $ meV) 
compared to 0.19 eV which is the averaged energy difference between the ground-state structure
and the second stable structure of Ag$_n$ (2$\le$n$\le$12) \cite{fournier}. 
Furthermore, the polarizabilities and the HOMO-LUMO gaps (HLg) do not offer a clear 
picture for elucidating the structural minimum of the octamer. That is, it is well known that 
the Ag$_8$ cluster is a
closed-shell system and it was demonstrated experimentally
and theoretically that the closure of electronic shell manifest itself 
in particularly large HLg ( see \cite{zhao} and references therein),
consequently the HLg reported in Table~\ref{table1} are on 
the side of stabilization of the T$_d$-TT isomer. 
As far as the reported polarizabilities are concerned, in molecular electronic distribution studies under the influence of an external electric
field, the relevant quantities are the mean static polarizability $\bar{\alpha}=  
(\sum{_{i=1}^3}\alpha_{ii})/3$ and
the polarizability anisotropy $\Delta\alpha$, defined as: 
\begin{equation}
	\label{eq1}
\Delta\alpha=\sqrt{\frac{\sum\limits _{i,j=1,2\atop i<j}^{2,3}(
\alpha_{ii}-\alpha_{jj})^2+6\sum\limits _{i,j=1,2\atop i<j}^{2,3}\alpha_{ij}^2}{2}} 
\end{equation}
where $\alpha_{ij}=\partial(\mu_e)_i/
\partial E_j$
is the ij-component of the polarizability tensor under the action of an external electric 
field $E_j$. It is not a common procedure to express the polarizability anisotropy
such as it was defined in Eq.~(\ref{eq1}). The commonly-used definition omits the 
second term ($6\sum_{i,j=1,2;i<j}^{2,3}\alpha_{ij}^2$) and thus 
neglects the important influence that the off-diagonal elements of the second-rank polarizability tensor 
play in the
symmetry considerations of the electric charge distribution \cite{alvarado}. 
The mean static polarizabilities, reported in Table~\ref{table1}, 
are quite similar to each other showing in average that the electron
charge is nearly equally distributed among the three isomers. However, only 
the polarizability anisotropy of the T$_d$-TT isomer is clearly reduced. This result
tends to stabilize the T$_d$ symmetry over D$_{2d}$ because the less polarizability anisotropy is, 
the more spherically symmetric charge distribution is and the latter condition is favored by a closed-shell 
system like the silver octamer. Indeed, the delocalized molecular orbitals for the higher-lying
occupied levels of the D$_{2d}$ and T$_d$ geometries, presented in Fig.~\ref{fig1}, exhibit a
hybridization of the atomic 5s levels with the 5p levels leading to a nearly spherical shape
whereas for the lower-lying unoccupied levels, the spherical symmetry becomes less important. The
proximity in energy and the corresponding superposition of the spatial distribution of the higher-lying 
occupied molecular orbitals of the T$_d$-TT isomer
with respect to the ones of the D$_{2d}$-DD isomer, contribute to a spherical symmetrization of the 
T$_d$-TT charge distribution.

\begin{figure}
\includegraphics[width=8.5cm,angle=0]{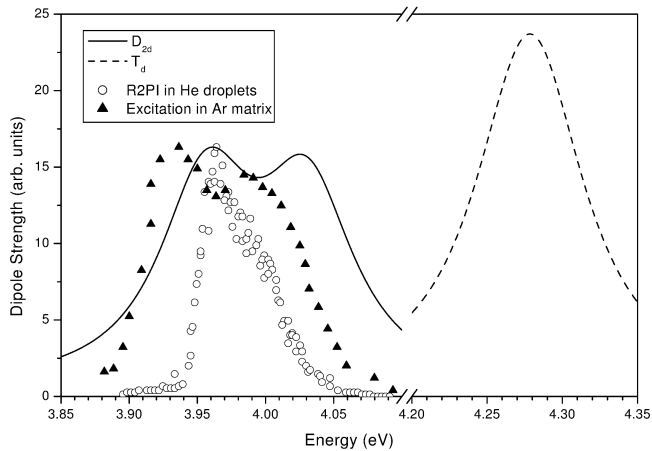}
\caption{\label{fig2}Comparison between two experimental recorded spectra  
and the calculated 
spectra for both (solid line) D$_{2d}$-DD and (dashed line) T$_d$-TT isomers
at a temperature of 10 K. 
Open circles 
correspond to the resonant two-photon-ionization (R2PI) spectroscopy on Ag$_8$ clusters in 
He droplets \cite{federmann} while the solid triangles are for the excitation spectrum of
Ag$_8$ excited with monochromatic Xe light in an Ar matrix \cite{felix}.
The D$_{2d}$-DD isomer spectrum is in excellent agreement with
the experiment whereas the T$_d$-TT isomer spectrum is around 0.31 eV blue-shifted compared to
the R2PI spectrum.}
\end{figure}

The balance of the aforementioned contrary tendencies does not allow a reliable
prediction of the structure, such as it was discussed in literature \cite{fournier,bonacic,erkoc,zhao,huda}.
The TDDFT calculations using the structural parameters provided by the LCGTO-KSDFM as starting point
can shed some light on the better understanding of this vexing controversy on the structure of Ag$_8$. 
In this respect, as it is shown in Fig.~\ref{fig2}, the calculated
spectrum for D$_{2d}$ symmetry is in excellent agreement with the resonant two-photon ionization
spectrum reported by F. Federmann {\it et al.} \cite{federmann} and the excitation spectrum reported
by C. F\'elix {\it et al.} \cite{felix}, whereas the T$_d$-TT calculated spectrum only shows one
resonant peak and it is about 0.31 eV 
blue-shifted with respect to the experimental measurements. Thus, the LCGTO-KSDFM predicted structure
is confirmed by the TDDFT calculations of the optical response when it is compared to the experimental 
evidences. 

Some of the relevant states involved in the transitions that populate the peaks of Fig.~\ref{fig2}
are provided through the calculated DOS depicted in Fig.~\ref{fig1}. For the D$_{2d}$ isomer, 
the eigenvalues of the
HOMO and HOMO-1 states are both close together in energy and the HOMO-2 state is 0.7 eV 
further down meanwhile for the T$_d$-TT isomer, this three states were grouped together in a window energy of only 3 meV. It is worthwhile to mention here that
the T$_d$-TT isomer has high symmetry, so many states will be degenerate.
However, because of small numerical errors, it is quite possible that states
that should  strictly be degenerate will show as being within very small
energy windows.
In the energy range displayed in Fig.~\ref{fig2} for D$_{2d}$ symmetry, the TDDFT calculations predict four
electronic transitions starting from the twofold degenerate HOMO and HOMO-1 states 
to two excited states separated 0.07 eV in energy and it gives rise to two peaks because of the nearness 
in energy of the HOMO-1 and HOMO states. In the case of the T$_d$ symmetry, six transitions are predicted
giving rise to only one peak because the two excited states that electronic transitions populate are only
1 meV separated, as commented above.

Some consequences can be extracted going further in the analysis of the calculated spectrum
for D$_{2d}$-DD isomer when it is compared with the experimental references reported in Fig.~\ref{fig2}.
On one hand, for the case of R2PI experiment we attribute the slight difference 
in peak position ($\sim$ 3 meV) to the helium environment through the formation of electron bubble states
that significantly blue shift the transition \cite{bartelt}. As already mentioned above, our TDDFT 
calculations also confirm the
authors' feeling of Ref.~\cite{federmann} that the asymmetry of the peak involves more than one 
transition. On the other hand, the comparison between the excitation spectrum in Ar matrix \cite{felix} and
the calculated D$_{2d}$ isomer spectrum allow us to measure the interaction of the Ar matrix with
the Ag$_8$ cluster. Thus, the shift of energy probably due to Ar matrix effects is estimated to be about 25 meV that is great enough 
compared to the slight difference in peak position ($\sim$ 3 meV) attributed to the helium 
surrounding the Ag$_8$ clusters in R2PI experiment. Consequently, the use of liquid helium droplets
as a spectroscopic matrix has the advantage over argon matrix of providing an environment more suitable
for study the electronic excitations of small and free silver clusters. 

In conclusion, we have shown that a combination of a LCGTO-KSDFM and TDDFT approach is able to reproduce the
measured optical response of the silver octamer and allow us to elucidate its lowest energy structure
below 10 K.
Our calculation thus confirms that the structural minimum of Ag$_8$ is the D$_{2d}$-DD isomer, whose 
geometrical structure is depicted in Fig.~\ref{fig1}. The TDDFT calculations have provided a number
of electronic transitions involving the resonant peaks showed in Fig.~\ref{fig2} and demonstrate that
the R2PI experiment is a good technique to measure experimentally the electronic excitations
of bare silver clusters because it is less aggressive than, for example,  the experiments that consider argon
as the spectroscopic matrix.

The authors acknowledge the CESGA for the computing facilities.
The work was supported by the Xunta de Galicia under
the Project No. PGIDIT02TMT20601PR.

\end{document}